\documentclass[conference,10pt,a4paper,twocolumn,twoside]{IEEEtran}
\IEEEoverridecommandlockouts

\usepackage[active]{srcltx} 
\usepackage{graphicx}
\usepackage{balance}
\usepackage{cite}
\usepackage{amssymb}
\usepackage[cmex10]{amsmath}
\usepackage{amsfonts}
\usepackage{amsthm}

\usepackage{eurosym}

\usepackage{hyperref} 

\usepackage{braket}
\usepackage{tikz}
\usetikzlibrary{arrows.meta}
\usepackage{booktabs}
\usepackage{subcaption}

\usepackage{algorithm}
\usepackage{algorithmicx}
\usepackage{algpseudocode}

\usepackage{mdframed}

\usepackage{siunitx}

\begin{document}

\newcommand{\eqdef}{\stackrel{\triangle}{=}}

% Definizione dei comandi
\newtheorem{theo}{Theorem}
\newtheorem{theor}{Theorem}
\newtheorem{cor}{Corollary}
\newtheorem{lem}{Lemma}
\newtheorem{prop}{Proposition}
\newtheorem{ins}{Insight}

\theoremstyle{remark}

\theoremstyle{definition}
\newtheorem{defin}{Definition}
\newtheorem{ass}{Assumption}
\newtheorem{rem}{Remark}
\newtheorem*{prob}{Optimal Quantum Routing Problem}

\renewcommand{\qed}{$\blacksquare$}

\renewcommand{\algorithmiccomment}[1]{// #1}

\renewenvironment{IEEEbiography}[1]
 {\IEEEbiographynophoto{#1}}
 {\endIEEEbiographynophoto}

\title{The Quantum Internet: Networking Challenges in Distributed Quantum Computing}
\author{
\IEEEauthorblockN{Angela Sara Cacciapuoti and Marcello Caleffi}
\IEEEauthorblockA{Dept. of Elect. Engineering and Inform. Technologies,\\
University of Naples Federico II, Naples, Italy \\
Email: \{angelasara.cacciapuoti, marcello.caleffi\}@unina.it}
\\   %<------ Line breaks in the current column
\IEEEauthorblockN{Francesco Saverio Cataliotti and Stefano Gherardini}
\IEEEauthorblockA{Dept. of Physics and Astronomy,\\
European Laboratory for Non-Linear Spectroscopy (LENS),\\
University of Florence, Florence, Italy\\
Email: \{fsc, gherardini\}@lens.unifi.it}
\and
\IEEEauthorblockN{Francesco Tafuri}
\IEEEauthorblockA{Dept. of Physics "Ettore Pancini",\\
University of Naples Federico II, Naples, Italy\\
Email: francesco.tafuri@unina.it}
\\  %<------- Extra vertical space
\IEEEauthorblockN{Giuseppe Bianchi}
\IEEEauthorblockA{Internet Engineering Department,\\
University of Roma Tor Vergata, Rome, Italy\\
E-mail:giuseppe.bianchi@uniroma2.it\\
\hphantom{linea invisibile}}
\thanks{\scriptsize{This work has been submitted to the IEEE for possible publication. Copyright may be transferred without notice, after which this version may no longer be accessible.}}
}

\maketitle

% --------------------------------------------------------------------------------------------------------------------------------------------------------------------------
% Abstract and keywords
% --------------------------------------------------------------------------------------------------------------------------------------------------------------------------
\begin{abstract}
The Quantum Internet is envisioned as the final stage of the quantum revolution, opening fundamentally new communications and computing capabilities, including the distributed quantum computing.
But the Quantum Internet is governed by the laws of quantum mechanics. Phenomena with no counterpart in classical networks, such as no-cloning, quantum measurement, entanglement and teleporting, impose very challenging constraints for the network design. Specifically, classical network functionalities, ranging from error-control mechanisms to overhead-control strategies, are based on the assumption that classical information can be safely read and copied. But this assumption does not hold in the Quantum Internet. As a consequence, the design of the Quantum Internet requires a major network-paradigm shift to harness the quantum mechanics specificities. The goal of this work is to shed light on the challenges and the open problems of the Quantum Internet design. To this aim, we first introduce some basic knowledge of quantum mechanics, needed to understand the differences between a classical and a quantum network. Then, we introduce quantum teleportation as the key strategy for transmitting quantum information without physically transferring the particle that stores the quantum information or violating the principles of the quantum mechanics. Finally, the key research challenges to design quantum communication networks are described.
\end{abstract}

\begin{IEEEkeywords}
Quantum Internet, Quantum Network, Entanglement, Teleporting.
\end{IEEEkeywords}

% --------------------------------------------------------------------------------------------------------------------------------------------------------------------------
% Section 1
% --------------------------------------------------------------------------------------------------------------------------------------------------------------------------
\section{Introduction}
\label{sec:1}
 
Nowadays, the development of quantum computers is experiencing a major boost, since tech giants entered the quantum race. In November 2017 IBM built and tested a 50-qubits processor, in March 2018 Google announced a 72-qubits processor, and other big players, like Intel and Alibaba, are actively working on double-digit-qubits proof-of-concepts. Meanwhile, in April 2017 the European Commission launched a ten-years 1~\euro -billion flagship project to boost European quantum technologies research. And in June 2017 China successfully tested a 1200 km quantum link between satellite Micius and ground stations.

Such a race in building quantum computers is not surprising, given their potential to completely change markets and industries - such as commerce, intelligence, military affairs \cite{CalCacBia-18,PirBra-16,WehElkHan18,Kim-08}. In fact, a quantum computer can tackle classes of problems that choke conventional machines, such as molecular and chemical reaction simulations, optimization in manufacturing and supply chains, financial modelling, machine learning and enhanced security.

The building block of a quantum computer is the quantum bit (qubit), describing a discrete two-level quantum state as detailed in the next section. By oversimplifying, the computing power of a quantum computer scales exponentially with the number of qubits that can be embedded and interconnected within \cite{Bou17,CalCacBia-18,Kim-08}. The greater is the number of qubits, the harder is the problem that can be solved by a quantum computer. For instance, solving some fundamental chemistry problems is expected to require\footnote{Qubit redundancy is mandatory due to the unavoidable presence of an external noisy environment.} ``hundreds of thousands or millions of interconnected qubits, in order to correct errors that arise from noise'' \cite{Bou17}. 
%
% %
%  \begin{figure}[t]
% 	\centering
% 	\includegraphics[width=.8\columnwidth]{./Fig-arch.png}
% 	\caption{Quantum Internet interconnects remote quantum devices to share quantum information.}
% 	\label{fig_arch}
% \end{figure}
% %
%
Quantum technologies are still far away from this ambitious goal. In fact, so far, although the quantum chips storing the qubits are quite small, with dimensions comparable to classical chips, they usually require to be confined into specialized laboratories hosting the bulked equipment -- such as large near absolute-zero cooling systems -- necessary to preserve the coherence of the quantum states. And the challenges for controlling, interconnecting, and preserving the qubits get harder and harder as the number of qubits increases. Currently, the state-of-the-art of the quantum technologies limits this number to double digits (IBM 50-qubits and Google 72-qubits).

Hence, very recently, the \textit{Quantum Internet} has been proposed as the key strategy to significantly scale up the number of qubits for long-distance communication of quantum and classical information \cite{CalCacBia-18,WehElkHan18,Kim-08}.

More in detail, the Quantum Internet is a \textit{quantum network}, i.e., a network able to connect remote quantum devices through quantum links in synergy with classical links (as detailed in Sec.~\ref{sec:3}). Such a quantum network constitutes a breakthrough, since it will provide unparalleled capabilities, as overviewed in \cite{WehElkHan18}, by exploiting its exponentially larger state space \cite{Kim-08}. Among them, there is certainly the \textit{Distributed Quantum Computing}. Specifically, by adopting the distributed paradigm, the Quantum Internet can be regarded as a virtual quantum machine constituted by a high number of qubits, scaling with the number of interconnected devices. This, in turn, implies the possibility of an exponential speed-up of the quantum computing power \cite{CalCacBia-18,YimLom-04}, with just a linear amount of the physical resources, i.e., the connected quantum devices.

However, from a communication engineering perspective, the design of the Quantum Internet is not an easy task at all. In fact, it is governed by the laws of quantum mechanics, thus phenomena with no counterpart in classical networks -- such as no-cloning, quantum measurement, entanglement and teleporting -- would impose terrific constraints to the network design. As instance, classical network functionalities, such as error-control mechanisms (e.g., ARQ) or overhead-control strategies (e.g., caching), are based on the assumption that classical information can be safely read and copied. But this assumption does not hold in a quantum network. As a consequence, the design of a quantum network requires a major paradigm shift to harness the key peculiarities of quantum information transmission, i.e., entanglement and teleportation. 

The goal of this paper is to shed light on the challenges in designing a quantum network. Specifically, we first introduce some basic knowledge of quantum mechanics -- superposition principle, no-cloning theorem, quantum measurement postulate and entanglement -- needed to understand the differences between a classical and a quantum network. Then, we focus on quantum teleportation as \textit{the key strategy} for transmitting quantum information without either the physical transfer of the particle storing the quantum information or the violation of the quantum mechanics principles. Finally, we draw the key research challenges related to the design of a quantum network. 

%We do not have the ambitious and arrogance of presenting an exhaustive analysis of such research topic, since it is at its infancy. However, we aim to contribute to such an exciting research area, which could pave the way for the Internet of future such as Arpanet paved the way for today's internet. }
% --------------------------------------------------------------------------------------------------------------------------------------------------------------------------
% Section 2
% --------------------------------------------------------------------------------------------------------------------------------------------------------------------------
\section{Quantum Mechanics Background}
\label{sec:2}
In this section, we briefly review some quantum mechanics postulates and principles needed to understand the challenges behind the design of a quantum network. 

A quantum bit, or \textit{qubit}, describes a discrete\footnote{At nano-scale, physical phenomena is described by discrete quantities, such as the energy levels of electrons in atoms or the horizontal/vertical polarization of a photon. This discrete nature contrasts with classical phenomena, which is commonly described by continuous quantities.} two-level quantum state, which can assume two (orthogonal) basis states: the zero (or ground state) and the one (or excited state), usually denoted\footnote{The \textit{bra-ket} notation $\ket{\cdot}$, introduced by Dirac, is a standard notion for describing quantum states. In a nutshell, a ket $\ket{\cdot}$ represents a column vector, hence the standard basis $\ket{0},\ket{1}$ is equivalent to a couple of 2-dimensional orthonormal vectors.} as $\ket{0}$ and $\ket{1}$. As instance, if we represent the state of a photon with a qubit, the two basis states represent the horizontal and the vertical polarization of the photon, respectively.

% --------------------------------------------------------------------------------------------------------------------------------------------------------------------------
\subsection{Superposition Principle}
\label{sec:2.1}
As widely known, a classical bit encodes one of two mutually exclusive states, being in only one state at any time. Conversely, a qubit can be in a superposition of the two basis states, so as to be simultaneously zero and one at a certain time. As an example, a photon with \ang{45} degrees of polarization is described by a superposed qubit, with an even \textit{amount} of zero and one, being simultaneously horizontally and vertically polarized.
Hence, while $n$ classical bits can encode only \textbf{one} of $2^n$ possible states at a certain time, $n$ qubits can simultaneously encode \textbf{all} the $2^n$ possible states at once, thanks to the superposition principle.

% --------------------------------------------------------------------------------------------------------------------------------------------------------------------------
\subsection{Quantum Measurement}
\label{sec:2.2}

According to one of the quantum mechanics postulates, whenever a measurement can have more than one outcome, as is the case for the two possible states of a qubit, after the measurement the original quantum state collapses in the measured state.
Hence, the measurement alters irreversibly the original qubit state \cite{NieChu-11}.
And, the result of such a measurement is probabilistic, since one obtains either the state zero or the state one, with a probability depending on the \textit{amount} of zero and one in the original superposed quantum state.
For instance, if the outcome of measuring a superposed qubit is the one corresponding to the state zero, the qubit collapses into such a state and any further measurement will give zero as outcome, independently of the original \textit{amount} of one in the superposed state.
As a consequence, although in principle a qubit can \textit{store} more than one classical bit of information thanks to the superposition principle, by measuring a qubit only one bit of information can be obtained.

The measurement postulate has deep implications on the quantum network design as described in Sec.~\ref{sec:4}. In fact, we can not ``share'' (i.e., transmit) quantum states among remote quantum devices by simply measuring the qubits and transmitting the measurement outcomes. To share qubits among remote devices without measuring them, we need to exploit a fundamental quantum mechanics property: \textit{the entanglement}.

% --------------------------------------------------------------------------------------------------------------------------------------------------------------------------
\subsection{No-cloning theorem}
\label{sec:2.3}
The no-cloning theorem states that an unknown qubit cannot be cloned, and it is a direct consequence of the quantum mechanics laws. Specifically, nature does not allow arbitrary transformations of a quantum system. Nature forces these transformations to be unitary. The linearity of unitary transformations alone implies the \textit{no-cloning theorem} \cite{NieChu-11}, which has critical consequences from a communication engineering perspective. In fact, classical communication functionalities are based on the assumption to be able to safely copy the information.
This in turn deeply affects the quantum network design, as pointed out in Sec.~\ref{sec:4}.

\begin{figure}[t]
	\centering
	\includegraphics[width=.75\columnwidth]{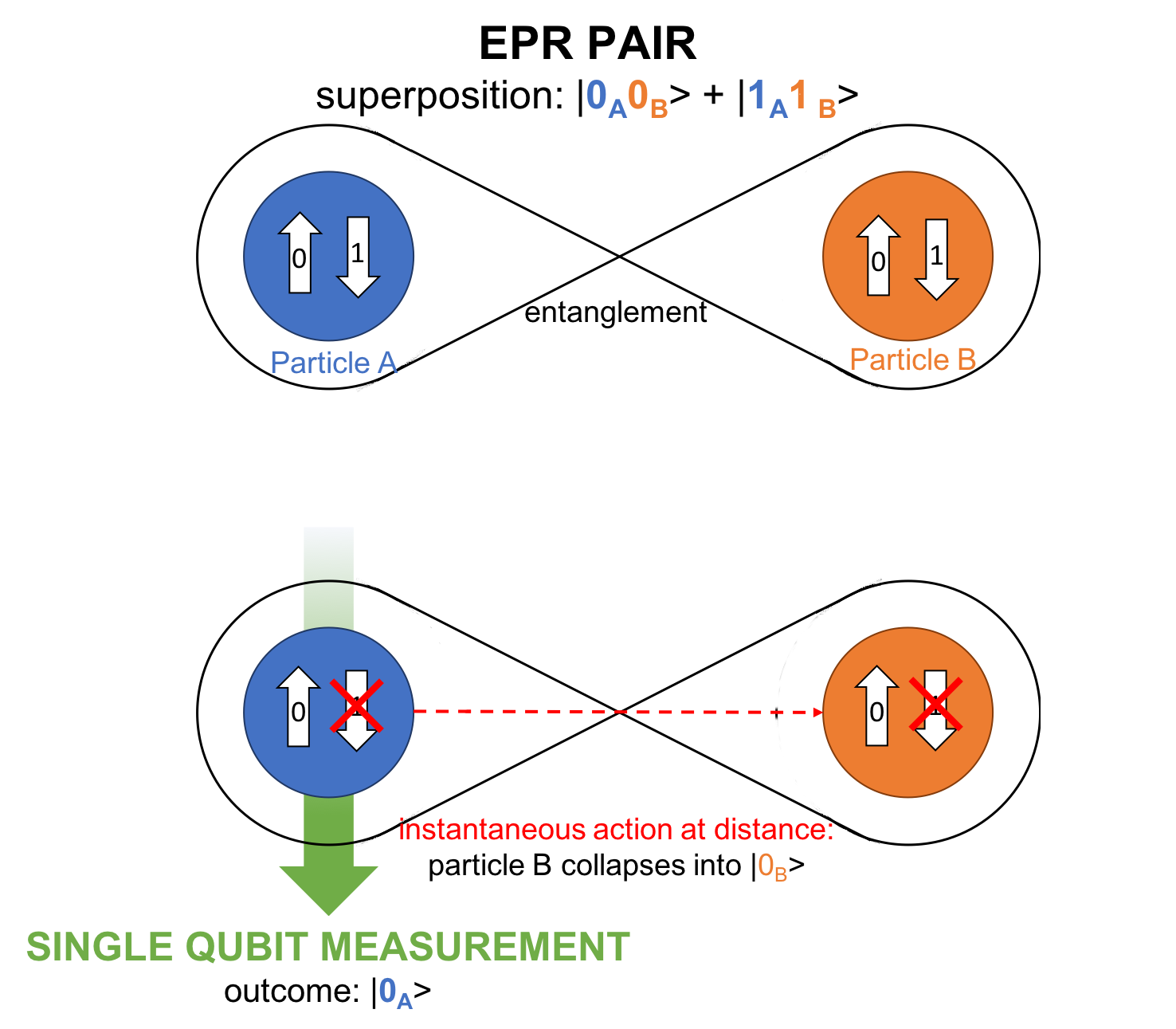}
	\caption{Quantum Entanglement: measuring one qubit of an EPR pair instantaneously changes the status of the second qubit, regardless of the distance dividing the two qubits. Specifically, the two particles forming the EPR pair are in superposed state, with an even \textit{amount} of zero and one. Hence, by measuring one of the two qubits, we obtain either zero or one with even probability. But once the qubit is measured, say particle A with outcome corresponding to state zero, the second qubit instantaneously collapses into state zero as well.}
	\label{fig:09}
\end{figure}

% --------------------------------------------------------------------------------------------------------------------------------------------------------------------------
\subsection{Entanglement}
\label{sec:2.4}
The deepest difference between classical and quantum mechanics lays in the concept of \textit{quantum entanglement}, a sort of correlation with no counterpart in the classical world. In a nutshell, the entanglement is a special case of superposition of multiple qubits where the overall quantum state can not be described in terms (or as a tensor product) of the quantum states of the single qubits.

To better understand the entanglement concept, let us consider Figure~\ref{fig:09}, showing a couple of maximally entangled qubits, referred to as \textit{EPR pair}\footnote{A couple of maximally entangled qubits is called EPR pair in honor of an article written by Einstein, Podolsky, and Rosen in 1935.}.
The couple of qubits forming the EPR pair are in a superposed state, with an even amount of zero and one. By measuring each of the two qubits independently, one obtains a random distribution of zero and one outcomes with equal probability. However, if the results of the two independent measurements are compared, one observes that every time that the measurement of qubit A yielded zero so did the measurement of qubit B, and the same happened with the outcome one. Indeed, according to quantum mechanics, as soon as one of the two qubits is measured the state of the other is \textit{instantaneously} fixed.

This quantum entanglement behavior induced Einstein and his colleagues to the so-called \textit{EPR paradox}: the measurement of one qubit instantaneously changes the state of the second qubit, regardless of the distance dividing the two qubits. This seems to involve information being transmitted faster than light, violating so the Relativity Theory. But the paradox is illusory, since entanglement does not allow to transmit information faster than light, as shown in Sec.~\ref{sec:3.1}.

\begin{figure*}[!t]
	\begin{minipage}[c]{0.48\textwidth}
		\centering
		\includegraphics[width=0.85\columnwidth]{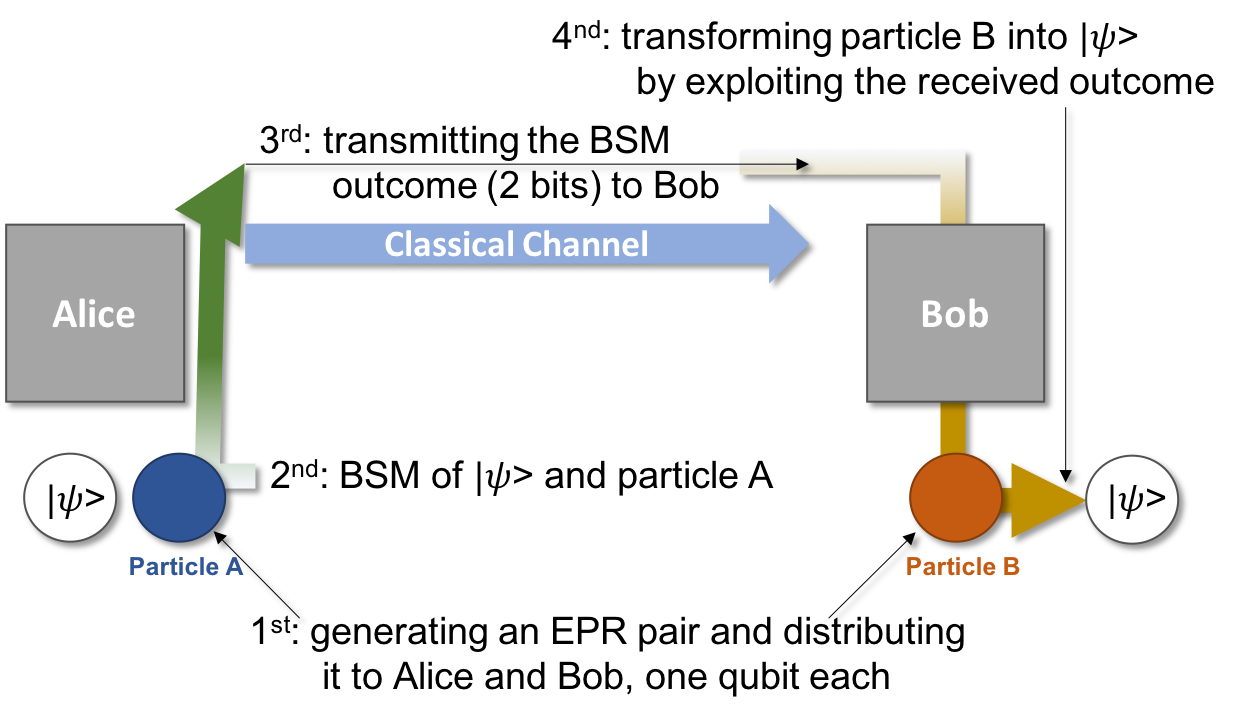}	
		\subcaption{High-Level Schematic. Two entangled qubits, forming an EPR pair, are generated and distributed so that one qubit (particle A) is stored by Alice and another qubit (particle B) is stored by Bob. Alice performs a BSM upon the two qubits at her side, i.e., the qubit $\ket{\psi}$ to be transmitted and particle A. Then, Alice sends the measurement outcome, i.e., 2 classical bits, to Bob with a classical channel. By processing particle B according to the measurement outcome, Bob finally obtains the qubit $\ket{\psi}$.}
		\label{fig:11.a}
	\end{minipage}
	\hspace{0.02\textwidth}
	\begin{minipage}[c]{0.48\textwidth}
		\centering
		\includegraphics[width=0.85\columnwidth]{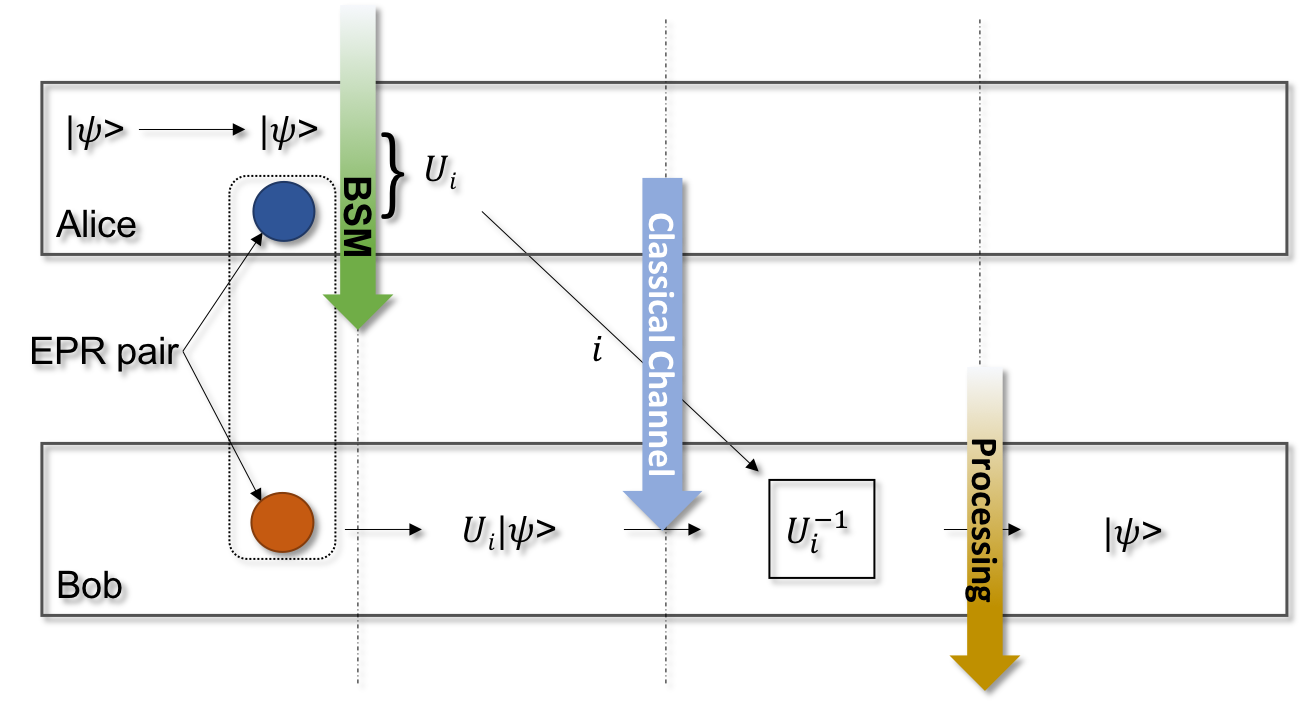}	
		\subcaption{Low-Level Schematic. An EPR pair is distributed among Alice and Bob so that particle A goes to Alice and particle B to Bob. Via a BSM upon particle A and the qubit $\ket{\psi}$ to be transmitted, the particle B at Bob's side collapses into a state resembling the qubit to be teleported, i.e. $U_i \ket{\psi}$. The result of Alice's measurement, sent through a classical channel, allows Bob to properly process particle B, i.e., to choose, among 4 possible operations, the unique operation $U_i^{-1}$ able to transform particle B into $\ket{\psi}$.}
		\label{fig:11.b}
	\end{minipage}
	\caption{Quantum Teleportation: unknown qubit $\ket{\psi}$ is ``transmitted'' from Alice to Bob by \textit{consuming} an EPR pair, shared between Alice and Bob. Relativity is not violated since $\ket{\psi}$ is obtained by Bob only after a classical communication between Alice and Bob occurred.}
	\label{fig:11}
\end{figure*}

% --------------------------------------------------------------------------------------------------------------------------------------------------------------------------

% --------------------------------------------------------------------------------------------------------------------------------------------------------------------------
% Section 3
% --------------------------------------------------------------------------------------------------------------------------------------------------------------------------
\section{Quantum Communications}
\label{sec:3}

\begin{table}[!b]
	\centering
	\caption{Classical vs Quantum}
	\includegraphics[width=1\columnwidth]{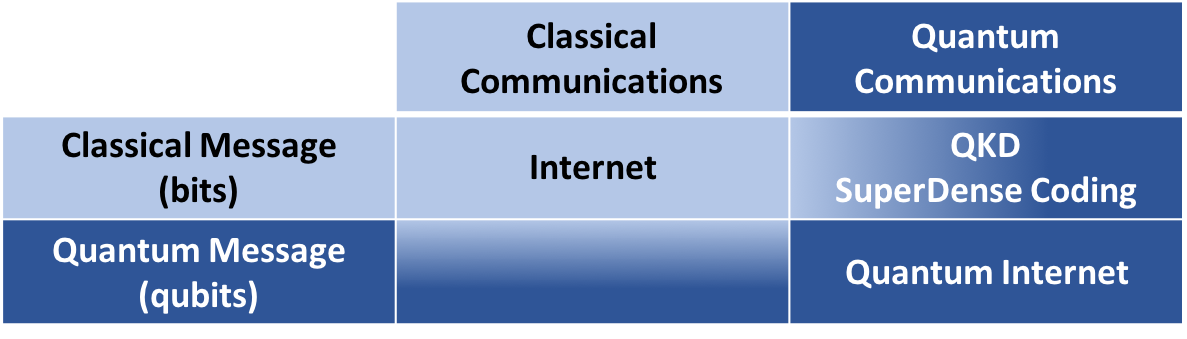}	
	\label{tab:01}
\end{table}

As widely known, classical communications, obeying to the laws of the classical physics, utilize bits to convey classical messages. Differently, quantum communications exploit quantum mechanics to fulfill the communications needs. However, so far, quantum communications have been widely restricted to a synonymous of specific applications, such as \textit{Quantum Key Distribution} (QKD) and \textit{superdense coding} \cite{NieChu-11}. Such applications exploit quantum mechanics only to convey classical messages (bits) as depicted in Table~\ref{tab:01}. Differently, the Quantum Internet expands and enriches the concept of quantum communications, since such a network can convey quantum messages (qubits), as depicted in Table~\ref{tab:01} as well.

More in detail, QKD is a cryptographic protocol enabling two parties to produce a shared random secret key by relying on the principles of quantum mechanics, either quantum measurement or entanglement. However, in a QKD system, quantum mechanics plays a role only during the creation of the encryption key: the encrypted information subsequently transmitted is entirely classical. Similarly, \textit{superdense} coding is a communication protocol enabling two parties to enhance the transmission of classical information through a quantum channel, i.e., to exchange two bits of classical information by exchanging a single qubit.

Differently, the Quantum Internet relies on the ability to share quantum states among remote nodes. However, quantum mechanics restricts a qubit from being copied or safely measured. Hence, although a photon can encode a qubit and it can be directly transmitted to a remote node, e.g., via a fiber link, if the traveling photon is lost due to attenuation or corrupted by noise, the quantum information is definitely destroyed. This quantum information can not be recovered via a measuring process and/or a copy of the original information, due to the postulate of quantum measurement and the no-cloning theorem. As a consequence, the direct transmissions of qubits via photons is not feasible and quantum teleportation, described in the following, must be employed.

% --------------------------------------------------------------------------------------------------------------------------------------------------------------------------
\subsection{Quantum Teleportation}
\label{sec:3.1}
Quantum Teleportation \cite{NieChu-11} provides an invaluable strategy for transmitting qubits without either the physical transfer of the particle storing the qubit or the violation of the quantum mechanics principles. Indeed, with just local operations, referred to as Bell\footnote{Named after J.S. Bell who, in 1964, proposed a physical test of the EPR paradox.}-State Measurement (BSM)\footnote{We emphasize that a solid strategy for a complete BSM with arbitrary photon states is not yet available.}, and an EPR pair shared between source and destination, quantum teleportation allows one to ``\textit{transmit}'' an unknown quantum state between two remote quantum devices. 

Differently from what the name seems to suggest, quantum teleportation implies the destruction of both the original qubit (encoding the quantum information to be transmitted) and the entanglement-pair member at the source, as a consequence of a measurement operation. Then, the original qubit is \textit{reconstructed} at the destination once the output of the BSM at the source -- 2 classical bits -- has been received at the destination through a classical channel.

The teleportation process of a single qubit is summarized in Figure~\ref{fig:11}. In a nutshell, it requires: i) the generation and the distribution of an EPR pair between the source and destination, ii) a classical communication channel to send the two classical bits resulting from the BSM measurement. 
Hence, it is worth noting that the integration of classical and quantum resources is a crucial issue for quantum networks\footnote{We note that, once Alice performed the measurement at her side, the qubit at Bob's side is instantaneously fixed. As a consequence, in principle, Bob can utilize his qubit (e.g., for a computational task) before receiving the two classical bits. And Bob can ``correct'' the result of its task \textit{a-posteriori}, whenever the two bits will be available at his side. This weak synchronization constraint -- which allows Bob to manipulate the qubit as long as he corrects the result (according to the two classical bits) before the end of the computational task -- can be leveraged in the communication protocols handling the integration between classical and quantum resources.}. 

Regarding the EPR pair, the measurement at the source side destroys the entanglement. Hence, if another qubit needs to be teleported, a new EPR pair must be created and distributed between the source and the destination. 

% --------------------------------------------------------------------------------------------------------------------------------------------------------------------------

% --------------------------------------------------------------------------------------------------------------------------------------------------------------------------
% Section 4
% --------------------------------------------------------------------------------------------------------------------------------------------------------------------------
\section{Quantum Network Design: Challenges and Open Problems}
\label{sec:4}

In this Section, we discuss the key research challenges and open problems related to the design of a quantum network, which harnesses quantum phenomena with no-counterpart in the classical reality, such as entanglement and teleportation, to share quantum states among remote quantum devices. 

% --------------------------------------------------------------------------------------------------------------------------------------------------------------------------
\subsection{Decoherence and Fidelity}
\label{sec:4.0}

Qubits are very fragile: any interaction of a qubit with the environment causes decoherence, i.e., a loss of information from the qubit to the environment as time passes.

Clearly, a perfectly-isolated qubit preserves its quantum state indefinitely. However, isolation is hard to achieve in practice given the current state-of-the-art of quantum technologies. Furthermore, perfect isolation is not desirable, since computation and communication require to interact with the qubits, e.g., for reading/writing operations. 

Decoherence is generally quantified through \textit{decoherence times}, whose values largely depend on the adopted technology for implementing qubits. For qubits realized with superconducting circuits\footnote{Superconducting qubits are based on Josephson tunneling junctions, which allow for a flow of electrons with a practically zero resistance between two superconductor layers. Under certain conditions, current can flow in both the directions, i.e., in a superposition of two states. As the two states are very close energetically, low temperatures (20 mK) are needed to avoid thermal fluctuations \cite{BenKel-15}.}, the decoherence times exhibit order of magnitude within $10$-$100 \,\mu$s \cite{BenKel-15}. Although a gradual decrease of the decoherence times is expected with the progress of the quantum technologies, the design of a quantum network must carefully account for the constraints imposed by the quantum decoherence.

Decoherence is not the only source of errors. Errors practically arise with any operation on a quantum state due to imperfections and random fluctuations. Here, a fundamental figure of merit is the \textit{quantum fidelity}. The fidelity is a measure of the distinguishability of two quantum states, taking values between 0 and 1. The larger is the imperfection of the physical implementation of an arbitrary quantum operation, the lower is the fidelity. Since teleportation consists of a sequence of operations on quantum states, the imperfection of such operations affects the fidelity of the ``transmitted" qubit.

From a communication engineering perspective, the joint modeling of the errors induced by the quantum operations, together with those induced by entanglement generation/distribution (as described in the next subsection), is still an open problem.

Furthermore, the no-cloning theorem prevents the adoption in quantum networks of classical error correction techniques, depending on information cloning, to preserve quantum information against decoherence and imperfect operations. Recently, many quantum error correction techniques have been proposed as in \cite{Hanzo18}. However, further research is needed. In fact, quantum error correction techniques must handle not only bit-flip errors, but also phase-flip errors, as well as simultaneous bit- and phase-flip errors. Differently, in the classical domain, a single type of error, i.e., the bit-flip error, has to be considered. 

From the above, the counteraction of the errors induced by decoherence and imperfect quantum operations in a quantum network is a functionality embracing aspects that traditionally belong to the physical layer of the classical network stack.

% --------------------------------------------------------------------------------------------------------------------------------------------------------------------------
\subsection{Entanglement Distribution}
\label{sec:4.1}

\begin{figure}[t!]
	\centering
	\includegraphics[width=0.8\columnwidth]{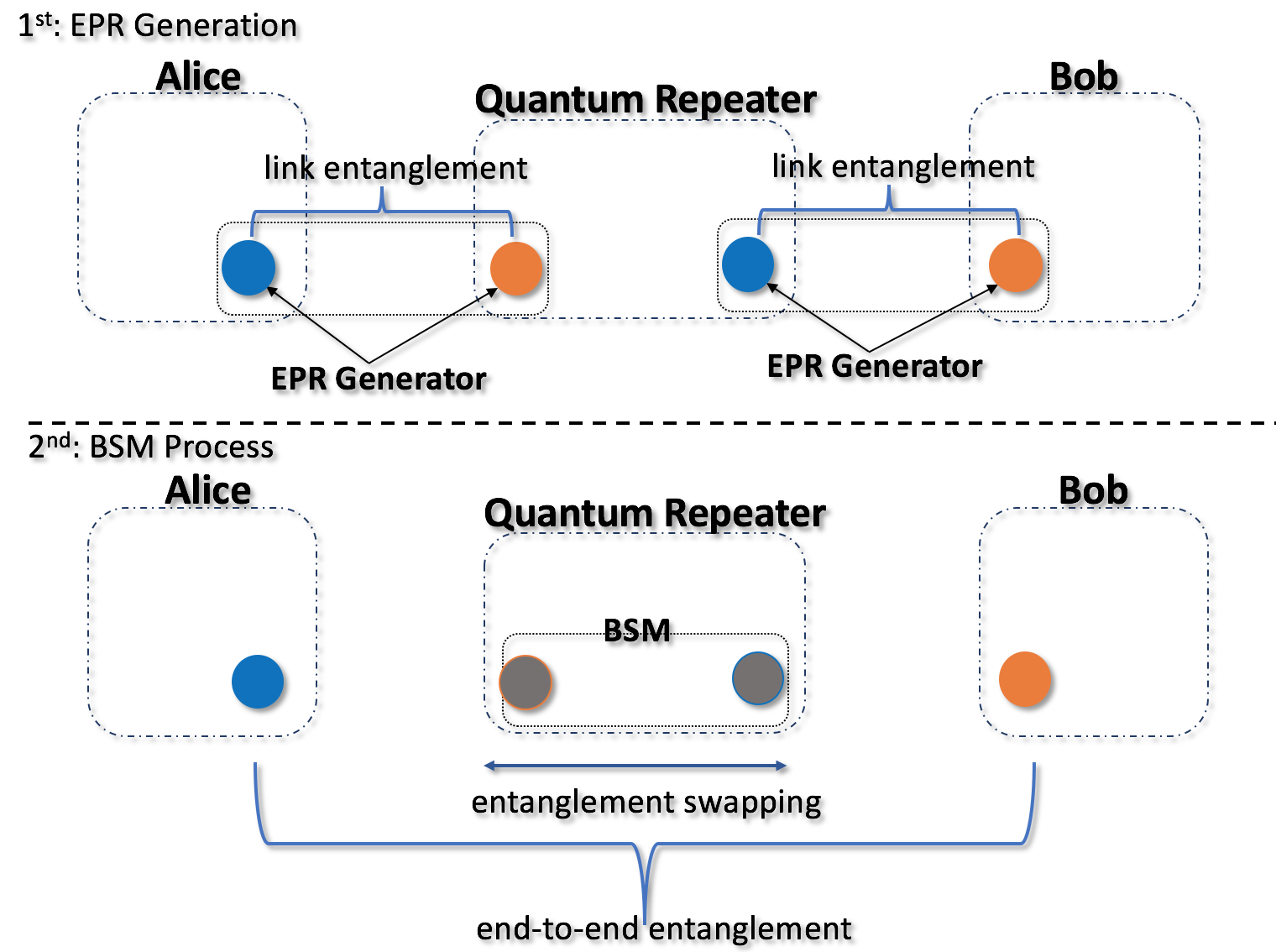}
	\caption{Entanglement Swapping. Two EPR pairs are generated and distributed: i) between a source (Alice) and an intermediate node (Quantum Repeater), and ii) between the intermediate node and a destination (Bob). By performing a BSM on the entangled particles at the Quantum Repeater, entanglement is eventually generated between Alice and Bob.}
    %Two EPR pairs are generated and distributed among a source (Alice), a destination (Bob) and an intermediate node (Quantum Repeater). By performing a BSM on the entangled particles at the Quantum Repeater, entanglement is created between Alice and Bob.}
	\label{fig:12}
\end{figure}

According to the quantum teleportation protocol, as in classical communication networks, the ``transmission'' of quantum information is limited by the classical bit throughput, necessary to transmit the output of the BSM outcome. But, differently from classical networks, the ``transmission'' of quantum information is achievable only if an EPR pair can be distributed between remote nodes.

Long-distance entanglement distribution -- although
deeply investigated by the physics community in the
last twenty years -- still constitutes a key issue due to the decay of the entanglement distribution rate as a function of the distance.

\begin{figure*}
	\centering
		\includegraphics[width=.75\textwidth]{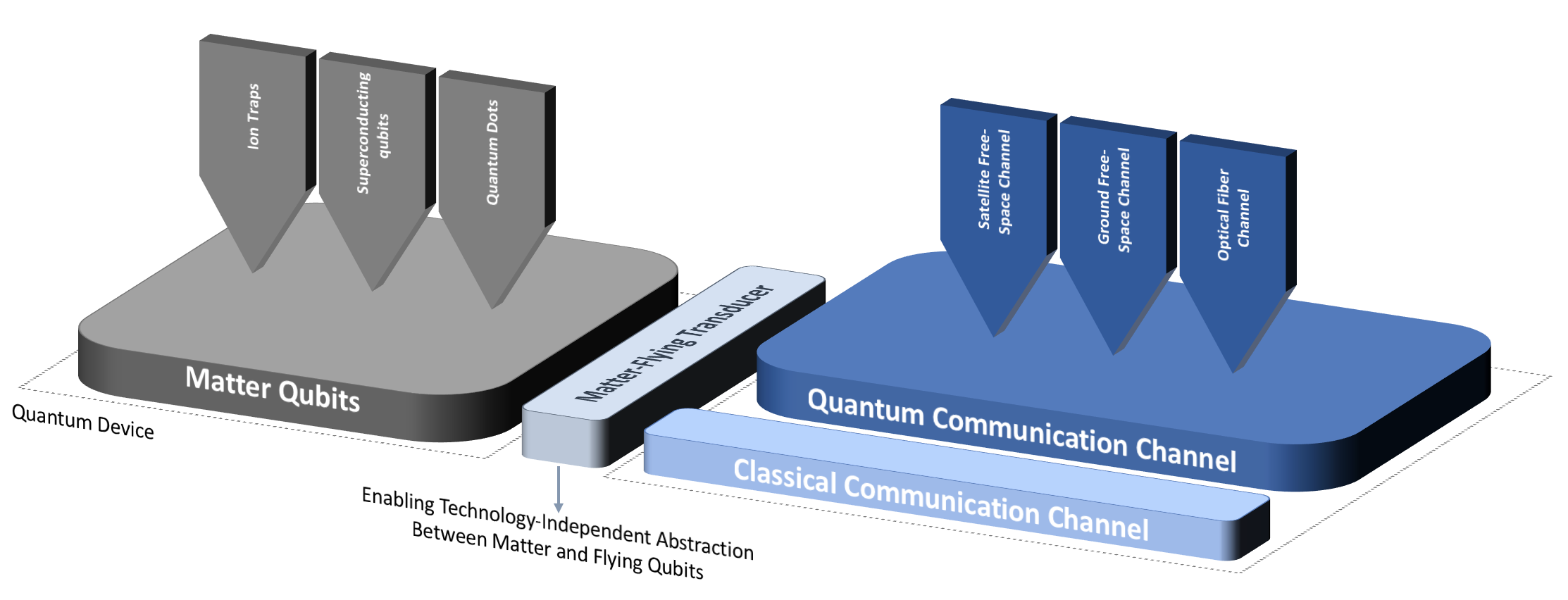}
	\caption{Pictorial representation of a Matter-Flying Interface for `transporting'' the qubit, encoding the quantum information to be transmitted, out of the physical quantum device at the sender side towards the corresponding quantum device at the receiver.}
	\label{fig:07}
\end{figure*}

There is a global consensus toward the use of photons as entanglement carriers, i.e., as candidates to generate entangled pairs among remote devices. Figure~\ref{fig:12} reports a possible strategy for entanglement distribution and we refer the reader to \cite{PirEisWee-15} for a survey. Entanglement distribution is achieved through Quantum Repeaters \cite{VanTou-13}, i.e. devices that implement a physical process known as \textit{entanglement swapping} \cite{VanTou-13}. In practice, two EPR pairs are generated with the source (Alice in Figure~\ref{fig:12}) and destination (Bob in Figure~\ref{fig:12}) receiving one element of each pair while the other two are sent to an intermediate node (the Quantum Repeater in Figure~\ref{fig:12}). By performing a BSM on the two entangled particles at the intermediate node, entanglement is created between the elements at the remote nodes. The process does not require the elements of the two pairs to have the same nature, therefore it can be used to distribute entanglement between two remotely located atoms or ions or to enhance the useful distance for the distribution of entangled photons. Hence, instead of distributing the entanglement over a long link, the entanglement is distributed iteratively through smaller links. In this regard, decoherence effects on the entanglement generation/distribution process can be mitigated via entanglement distillation, which can be regarded as a type of error-correction for quantum communication between two parties \cite{NieChu-11}. The distillation procedure, however, requires additional levels of qubit processing.
Despite these efforts, further research is needed, especially from an engineering perspective. In fact, the entanglement distribution is a key functionality of a quantum network embracing aspects belonging to different layers of the classical network stack. More in detail:
\begin{itemize}
\item \textit{Physical Layer}: The entanglement is a perishable resource. Indeed, due to the inevitable interactions with the external environment, the entanglement among entangled parties is progressively lost over time, and entanglement losses deeply affect the quality of the teleportation process. For the reasons highlighted in the previous subsection, classical error correction techniques cannot be used to counteract these losses. Hence, robust entanglement distribution techniques based on quantum error correction are mandatory for the deployment of a quantum network. This still represents a very hard challenge in this field.

\item \textit{Link Layer}: The no-broadcasting theorem, a corollary of the no-cloning theorem, prevents quantum information from being transmitted to more than a single destination. This is a fundamental difference with respect to classical networks, where broadcasting is widely exploited for implementing several layer-2 and -3 functionalities, such as medium access control and route discovery. As a consequence, the link layer must be carefully re-thought and re-designed, and effective multiplexing techniques for quantum networks should be designed to allow multiple quantum devices to be connected to a single quantum channel (e.g. a fiber). The access to the medium could be based for example on photon-frequency-division for the entanglement distribution.

\item \textit{Network Layer}: The entanglement distribution determines the connectivity of a quantum network in term of capability to perform teleporting among the quantum devices. Hence, novel quantum routing metrics are needed to ensure effective entanglement-aware path selection. Furthermore, the teleportation process destroys the entanglement as a consequence of the BSM at the source. Hence, if another qubit needs to be teleported, a new entangled pair needs to be created and distributed between the source and the destination. This constraint has no-counterpart in classical networks and it must be carefully accounted for in an effective design of the network layer.
\end{itemize}

% --------------------------------------------------------------------------------------------------------------------------------------------------------------------------
\subsection{Interface between Matter Qubits and Flying Qubits}
\label{sec:4.2}

As mentioned before, there exists a general consensus about the adoption of photons as substrate for the so-called \textit{flying} qubits, i.e., as entanglement carriers. The rationale for this choice lays in the advantages provided by photons for entanglement distribution: weak interaction with the environment (thus, reduced decoherence), easy control with standard optical components as well as high-speed low-loss transmissions.
The aim of the flying qubits is to ``transport'' qubits out of the physical quantum devices through the network for conveying quantum information from the sender to the receiver. Hence, as shown in Figure~\ref{fig:07}, a transducer \cite{Ham10} is needed to convert a \textit{matter} qubit, i.e., a qubit for information processing/storing within a computing device, in a flying qubit, which creates entanglement among remote nodes of the network. 

Nowadays, there exist multiple technologies for realizing a matter qubit (quantum dots, transmon, ion traps, etc) and each technology is characterized by different pros and cons, as surveyed in \cite{VanDev-16}. As a consequence, a matter-flying interface is required also to face with this technology diversity.

Moreover, from a communication engineering perspective, the interface should be compatible also with the peculiarities of the physical channels the flying qubits propagate through. In fact, there exist different physical channels for transmitting flying qubits, ranging from free-space optical channels (either ground or satellite free-space) to optical fibers. 

In the last ten years, the physics community has been quite active on investigating schemes and technologies enabling such an interface, with an heterogeneity of solutions \cite{Ham10,StaRabSor-10} ranging from cavities for atomic qubits to optomechanical transducers for superconducting qubits. 
As a consequence, the communication engineering community should join these efforts by designing communication models that account both the technology diversity in fabricating qubits and the propagation diversity in characterizing the different physical channels. 

% -------------------------------------------------------------------------------------------------------------------------------------------------------------------------
\subsection{Deployment Challenges}
\label{sec:4.3}

Quantum Internet is probably still a concept far from a real world implementation, but it is possible to outline a few deployment challenges for the near future.
\begin{itemize}
	\item The current technological limits to qubits and quantum processors physical realizations:\\
    At first, quantum computers will be available in few, highly specialized, data centers capable of providing the challenging equipment needed for quantum computers (ultra-high vacuum systems or ultra low temperature cryostats). Companies and users will be able to access quantum computing power as a service via cloud. In this regard, the quantum cloud market is estimated nearly half of the whole 10 billion quantum computing market by 2024 \cite{HSRC-18}. IBM already allows researchers to practice quantum algorithm design through a classical cloud access to isolated 5-, 16- and 20-qubits quantum devices.
	\item Existing technological limits to quantum communication and quantum interfaces:\\
	The first realizations of a Quantum Internet will be small clusters of quantum processors within a data center. Architectures will have to take into account the high cost of data buses (economically and in term of quantum fidelity) limiting both the size of the clusters and the use of connections for processing.
\item Hybrid architectures will probably be used for connections faring both the use of cryo-cables (expensive and necessarily limited in length) and of optical fibers or free space photonic links.
\item Given the fragile nature of quantum entanglement and the challenges posed by the sharing of quantum resources, a substantial amount of conceptual work will be needed in the development both of novel networking protocols and of quantum and classical algorithms.
\item As described in Sec.~\ref{sec:3.1}, quantum teleportation requires the integration of classical and quantum communication resources. The classical communication resources will be likely provided by integrating classical networks such as the current Internet with the Quantum Internet. This represents a completely unexplored open problem, and its solution requires a multidisciplinary effort, spanning the breath from communications theory and engineering communities, to the networking engineering one.
\end{itemize}

In conclusion, the Quantum Internet, though still in its infancy, is a very interesting new concept where a whole new set of novel ideas and tools at the border between quantum physics, computer and telecommunications engineering will be needed for the successful development of the field.

%\balance 

% ----------------------------------------------------------------------------------------------------------------------------------------------------------------------
% Bibliography
% ----------------------------------------------------------------------------------------------------------------------------------------------------------------------
\bibliographystyle{IEEEtran}
\bibliography{quantum-04}
%\balance 

\vspace{-13 mm}

\begin{IEEEbiography}
{Angela Sara Cacciapuoti} (M'10-SM'16) is a Tenure-Track Assistant Professor at the University of Naples Federico II, Italy. Since July 2018, she held the national habilitation as "Full Professor" in Telecommunications Engineering. Her work appeared in first tier IEEE journals, and she received several awards. Currently, Angela Sara serves as Editor/Associate Editor for: IEEE Trans. on Communications, IEEE Communications Letters, and IEEE Access. In 2016 she was elevated to IEEE Senior Member and she has been an appointed member of the IEEE ComSoc YP Standing Committee. Since 2018, she is the Publicity Chair of the IEEE ComSoc WICE Standing Committee.
\end{IEEEbiography}

\vspace{-13 mm}

\begin{IEEEbiography}
{Marcello Caleffi} [M’12, SM’16] is a tenure-track assistant professor at the University of Naples Federico II. His work has appeared in several premier IEEE transactions and journals, and he has received multiple awards, including best strategy, most downloaded article, and most cited article awards. Currently, he serves as editor for IEEE Communications Magazine, IEEE Access, and IEEE Communications Letters. He serves as Distinguished Lecturer for the IEEE Computer Society and, since 2017, he serves as elected treasurer for IEEE ComSoc/VT Italy Chapter.
\end{IEEEbiography}

\vspace{-13 mm}

\begin{IEEEbiography}
{Francesco Tafuri} is a Full Professor at the University of Naples Federico II. His scientific interests fall in the field of Superconductivity. He is leading an experimental research group investigating macroscopic quantum phenomena and the Josephson effect in hybrid devices, new routes for the realization of superconducting qubits for hybrid architectures, mesoscopic and nanoscale systems, vortex physics. He is Associate Editor of Journal of Superconductivity, in the Advisory Board of Physica C, and Peer Reviewer of National and European projects. He is author of more than 160 publications on international journals and books, including 13 papers on Science/Nature/Nat.Mat./Nat.Comm./PRL. He has given $>$90 invited talks at international conferences and research institutes. Several manuscripts are the product of international and national collaborations, including IBM T.J. Watson Research Center, USA; Stanford University, USA; Chalmers University of Technology, Goteborg, Sweden; University of Cambridge, UK; Lawrence Berkeley Laboratory, USA; Columbia University, USA.
\end{IEEEbiography}

\vspace{-13 mm}

\begin{IEEEbiography}
{Francesco Saverio Cataliotti} activity is mainly experimental and is centered on atomic physics and atom-laser interactions. It can be divided in lines that, starting form nonlinear atom-laser interaction, develop into the manipulation of the atomic momentum state, laser-cooling and the realization of Bose-Einstein condensates. The optical manipulation of condensates has allowed him to investigate problems concerning superfluidity and macroscopic quantum states. More recently he has been concerned with coherent micro-manipulation techniques with atoms, molecules and micro-mechanical systems with perspectives ranging from quantum computation to single molecule manipulation. He is author of more than 100 scientific papers on international peer reviewed journals and contributions to conferences and workshops. Overall his works have received more than 2900 citations to date.
\end{IEEEbiography}

\vspace{-13 mm}

\begin{IEEEbiography}
{Stefano Gherardini} was born in 1989. He received his B.Sc. degree in Electronic and Telecommunications Engineering and his M.Sc. degree in Electrical and Automation Engineering from the University of Florence, Italy, in 2011 and 2014, respectively. He got his Ph.D. degree in Information Engineering, curriculum non-linear dynamics and complex systems, in 2018 with the Automatic Control and QDAB group at the University of Florence. Moreover, he is a visiting student at SISSA, the International School for Advanced Studies, in Trieste, Italy. His main research interests include binary measurements, networked moving-horizon estimation, Kuramoto models, disordered systems, stochastic quantum Zeno effect, open quantum systems theory, and quantum estimation and control.
\end{IEEEbiography}

\vspace{-15 mm}

\begin{IEEEbiography}
{Giuseppe Bianchi} has been a full professor of networking at the School of Engineering of the University of Roma “Tor Vergata” since 2007. His research activity, documented in more than 200 peer-reviewed international journal/conference papers having received to date more than 14,000 citations, focuses on wireless networks (his WLAN modeling work received the ACM SIGMO-BILE Test Of Time award in 2017), programmable network systems, and privacy and security monitoring in both the wireless and wired domains.
\end{IEEEbiography}

\end{document}